\documentclass[conference]{IEEEtran}
\IEEEoverridecommandlockouts
\usepackage{cite}
\usepackage{amsmath,amssymb,amsfonts}
\usepackage{algorithmic}
\usepackage{graphicx}
\usepackage{textcomp}
\usepackage{pgfplots}
\pgfplotsset{width=7cm,compat=1.8}
\usepackage{colortbl}
\usepackage{pgfplotstable}
\usepackage{xcolor}
\usepackage{booktabs}
\usepackage{tikz}
\usepackage{collcell}
\usepackage{diagbox}
\newcommand*{\MinNumber}{0.293}%
\newcommand*{\MidNumber}{0.672} %
\newcommand*{\MaxNumber}{0.998}%

\newcommand{\ApplyGradient}[1]{%
    \ifdim #1 pt > \MidNumber pt
    \pgfmathsetmacro{\PercentColor}{max(min(100.0*(#1 - \MidNumber)/(\MaxNumber-\MidNumber),100.0),0.00)} %
    \hspace{-0.33em}\colorbox{red!\PercentColor!yellow}{#1}
    \else
    \pgfmathsetmacro{\PercentColor}{max(min(100.0*(\MidNumber - #1)/(\MidNumber-\MinNumber),100.0),0.00)} %
    \hspace{-0.33em}\colorbox{green!\PercentColor!yellow}{#1}
    \fi
}

\newcolumntype{R}{>{\collectcell\ApplyGradient}c<{\endcollectcell}}

\def\BibTeX{{\rm B\kern-.05em{\sc i\kern-.025em b}\kern-.08em
T\kern-.1667em\lower.7ex\hbox{E}\kern-.125emX}}
\begin{document}

\title{Idiolect: A Reconfigurable Voice Coding Assistant}

\author{\IEEEauthorblockN{Breandan Considine}
\IEEEauthorblockA{\textit{McGill University} \\
bre@ndan.co}
\and
\IEEEauthorblockN{Nicholas Albion}
\IEEEauthorblockA{\textit{Independent Developer} \\
nalbion@yahoo.com}
\and
\IEEEauthorblockN{Xujie Si}
\IEEEauthorblockA{\textit{University of Toronto} \\
six@cs.utoronto.ca}
}

\maketitle

\begin{abstract}
    This paper presents Idiolect, an open source~\footnote{https://github.com/OpenASR/idiolect} IDE plugin for voice coding and a novel approach to building bots that allows for users to define custom commands on-the-fly. Unlike traditional chatbots, Idiolect does not pretend to be an omniscient virtual assistant but rather a reconfigurable voice programming system that empowers users to create their own commands and actions dynamically, without rebuilding or restarting the application. We offer an experience report describing the tool itself, illustrate some example use cases, and reflect on several lessons learned during the tool's development.
\end{abstract}


\section{Introduction}

Humans are able to quickly learn new words and phrases, and apply them in a variety of contexts. Chatbots, however, are often limited to a set of static commands and phrases. This creates a frustrating experience for users, who struggle to express their intent, as well as bot developers, who must anticipate user intent and make new capabilities discoverable. This rigidity is a common source of misalignment between user intent, author expectations and bot capabilities.

Voice coding allows users to quickly dictate new commands and behaviors. For example, one might say, ``whenever I say \textit{open sesame}, do the following action'', thereafter, the system will perform the desired action when so instructed. Or, ``whenever I say \textit{redo thrice}, repeat the last action three times''. Or, e.g., invoke a function in a scripting language, open a file, or perform other manually repetitive chores.

Idiolect provides a default lexicon of phrases, but does not impose them upon end-users. Instead, users may override the defaults with custom voice commands on-the-fly, which are incorporated into the lexicon without delay. By shifting the burden of adaptation from the user to the system, this frees our users to express their intent in a more natural manner.

Primarily, Idiolect observes the following design principles: be (1) natural to use, (2) easy to configure, (3) as unobtrusive as possible. We believe that these principles are important for a system intended to be used by developers, who are busy people and capable of configuring the system themselves. We also support developers with visual and motor impairments, who may have difficulty typing, or prefer to use a voice interface.

In the following paper, we describe Idiolect, an IDE plugin originally developed at JetBrains in 2015 and recently updated to provide many new features, such as deep speech recognition and dynamically-reconfigurable voice commands, letting users verbally or programmatically express their desired intent.

\section{Prior Work}

Mary Shaw, during her 2022 SPLASH keynote~\cite{shaw2022myths} called for programming languages to address the needs of ``vernacular developers''. Jin Guo has also talked about the need for programming in ``ordinary people's language''. We take their proposals quite literally to mean that computers should be able to interpret spoken programs, and not just written ones.

Although related, voice assistants like Siri and Alexa are not configurable nor intended for programming. In this regard, we believe our tool is more akin to a programmable bot with audio I/O than a dedicated voice assistant. Spoken programs are also closely related to natural language programming, an idea once scorned~\cite{dijkstra1979foolishness}, but now becoming increasingly plausible.

Early attempts to build voice programming systems can be traced back at least two decades to Leopold and Amber's~\cite{leopold1997keyboardless} (1997) work on keyboardless programming, later revisited by Arnold and Goldthwaite~\cite{arnold2000programming} (2001), Begel and Graham~\cite{begel2005spoken} (2005) and others. These systems allow users to write code by speaking into a microphone, however early voice programming systems were limited by a small vocabulary, and unlike our work, do not consider IDE integration or reconfigurability.

Another stream of work has explored teachable voice assistants and user-defined phrases (Chkroun \& Azaria~\cite{chkroun2019lia}). Though similar to our own, their work targets general-purpose voice user interfaces and does not directly consider voice programming. It also predates most of the recent progress on large language modeling, which we consider to be a transformative enabling technology for programming by voice.

\section{Architecture}

Today, automatic speech recognition (ASR), the translation of an audio waveform containing speech to text, is essentially a solved problem -- one of the many pretrained ASR models would work well enough for our purposes. However, we also require realtime offline speech recognition capabilities built on an open source deep speech pipeline, which has only recently become possible for users running on commodity hardware.

Idiolect integrates with Vosk, a state-of-the-art deep speech system with realtime models for various languages, which provides an open source Java API. This allows us to provide a high-quality speech recognition experience without requiring users to upload their personal data to a cloud service.



Users may optionally configure a built-in TTS voice from the host operating system and a cloud-based speech recognition or synthesis service, with the caveat that web speech requires uninterrupted internet connectivity and introduces an additional 300-500ms of overhead latency depending on the user's proximity to the datacenter and other load factors.

\begin{figure}[t]
    \centering
    \includegraphics[width=0.9\linewidth]{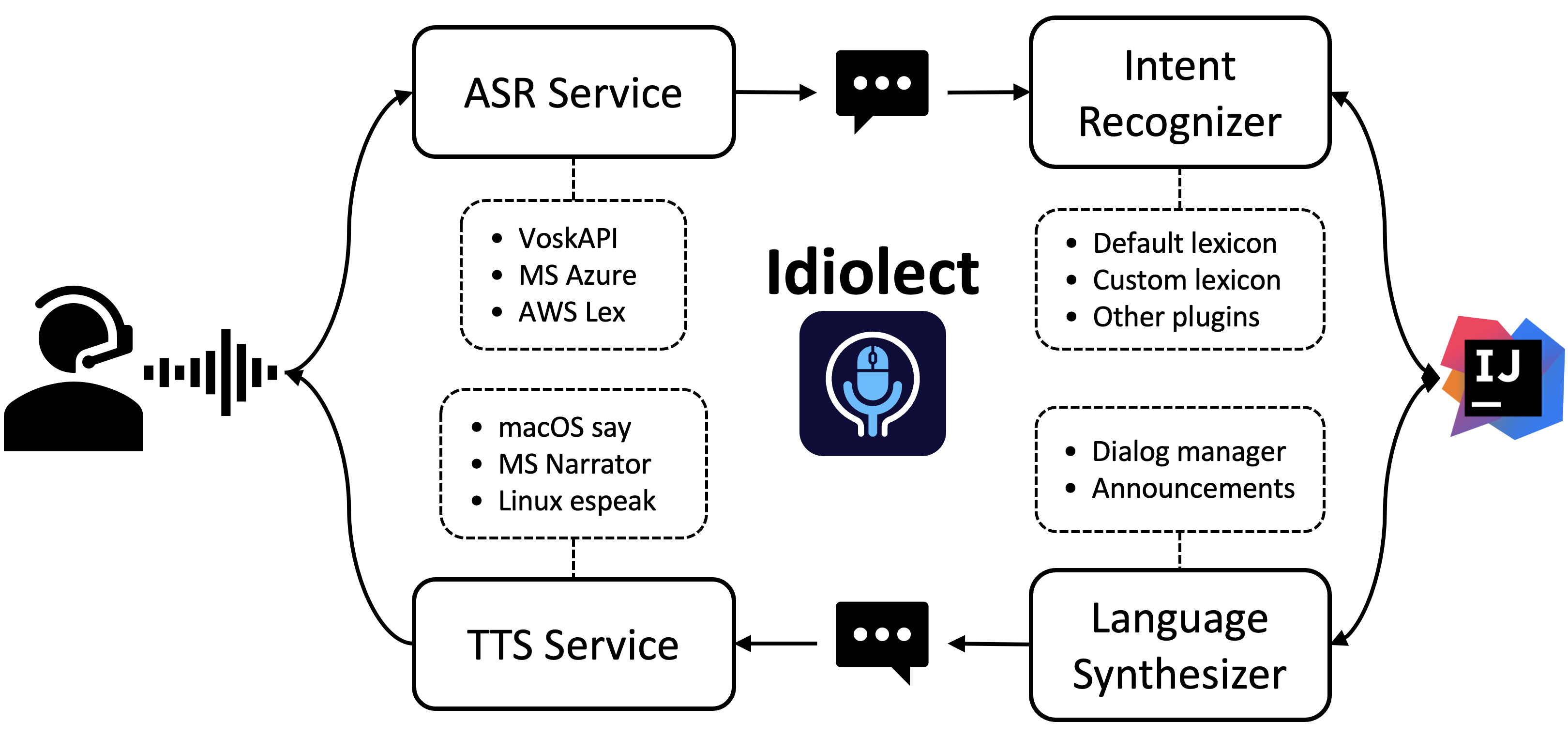}
    \caption{Idiolect has four components: a speech and intent recognizer, language synthesizer, and TTS service, which together faciliate a dialog with the IDE.}
    \label{fig:architectural_overview}
\end{figure}

Once a spoken utterance is decoded as text, Idiolect must determine the relevant actions and entities needed to fulfill the user's intent. Furthermore, it may need to consider the IDE context, i.e., the current editor state and command history, to resolve potentially ambiguous commands. For example, the command, ``open plugin menu'' could refer to multiple different menus, depending on when and how it was invoked.

The IntelliJ Platform has over $10^3$ possible actions. These actions are bound to keyboard shortcuts, menu items, and toolbar buttons. The user can also bind a voice command directly to an action, presuming the user already knows the action's identifier. Idiolect's default grammar was manually curated from the IDE action list, using the CamelCase identifier to generate a suitable description for intent recognition.

In keeping with the principle of configurability, we allow users to define custom grammars and bind utterances to actions using a context-free grammar. By providing an extensible DSL, users can bind their own command patterns using either a simple configuration file, or programmatically via the plugin API to handle more complex usage scenarios.



Idiolect dispatches utterances to a series of recognizers using a turn-based priority queue, in which each recognizer is given a single turn to match or pass on each utterance. Once a command is recognized, the command is consumed and dispatched to no subsequent recognizer thereafter, which prevents a single command from triggering multiple actions.

The plugin first attempts to resolve an utterance using an exact lexical match against a finite language, i.e., a lexicon of predefined commands. Highest priority are those which control the plugin itself, enabling and disabling speech recognition.

User-defined commands are the next highest priority. These can be matched lexically, or more generally as a sentence in a context-free language. Nonterminal parameters allow one to define more complex instructions, such as ``open the \textlangle filename\textrangle [in \textlangle project\_name\textrangle]'' or ``jump to the \textlangle nth\textrangle line''.


The recognizer of last resort is a large language model. Open-ended commands which are unrecognizable to all of the previous methods are dispatched to a service that matches the intent against a set of available actions using a prompt, e.g., ``Out of these actions: \ldots, which one most likely fulfills the command \ldots?'', whose response is then dynamically invoked.


\pagebreak\section{Barriers and Pathways to Usability}\label{sec:usability}

In our experience as plugin users and maintainers, usability challenges typically arise in a few key areas. In the following section, we will discuss some of those obstacles encountered while developing Idiolect and our efforts to overcome them.

\subsection{User Onboarding}

Several users reported confusion when first installing our plugin. To address this issue, we added a wizard that guides users through installation. Upon first installing the plugin, a user is greeted and prompted to download the Vosk model for recognizing their preferred natural language, defaulting to the system locale. Once unpacked, the model is stored in the plugin configuration directory, \texttt{~/.idiolect}. The user is next prompted to configure the properties file, and bind a few voice commands. This is a one-time process, and users may reconfigure the plugin via the settings menu at any time.

\subsection{Plugin Observability}

Failures arise in many stages of the intent recognition pipeline. Short of fixing the error directly, often the best way to address failure is by improving observability, so that users can diagnose an error themselves and avoid it in the future.

Recognition failure is a common issue, often caused by transcription errors in the ASR model, due to, e.g., noise in the audio signal, stopwords, or poor recognition accuracy. To address this issue, we add a visual cue at the corner of the IDE that reports the phrase transcribed in realtime, as well as the action (if any) that was triggered on intent recognition.

\subsection{Command Discoverability}

We draw a distinction between capability discovery and intent recognition. Users previously familiar with the IDE capabilities may wish to invoke or bind a specific action directly, but may not know the specific action identifier. The first and foremost way to improve discovery is through documentation, however keeping documentation in sync with capabilities can be challenging, and users are uninclined to read verbose documentation. To address this issue, we autogenerate documentation for each action by preprocessing the action list and annotating each action with a natural language description from the IntelliJ Platform source code.

Another common scenario is when the user is unfamiliar with the IDE capabilities, and the action they wish to perform is not supported by the IDE directly. User awareness of IDE functionality is outside the scope of this plugin, however, certain actions users may intend to perform are compound actions or have no associated binding. For example ``delete the method named \textlangle foo\textrangle'' requires first resolving the method, selecting the body of the text and deleting it. Such actions can be defined, but require implementing a handcrafted recognition handler, and are generally quite brittle. We hope to improve scripting support for such actions in the future.

\subsection{API Extensibility}

Idiolect tries to anticipate users actions and expose default action bindings. Some functionality, however, is best left implemeneted by downstream developers, who are better-equipped to understand their users' needs. In addition to end-user configurability, Idiolect is designed to be extensible by other IntelliJ Platform plugins, which can define additional commands, recognition handlers and custom actions. We offer a simple message-passing API for plugins to communicate with Idiolect, and a DSL for defining custom grammars.

\subsection{Error Recovery}\label{sec:error}

Another common usability barrier is when transcription is accurate, but the utterance does not correspond to an actionable command, possibly due to stopwords, verbal fillers or extraneous text which cannot be directly parsed. In short, if a given phrase, e.g., ``open uh foo java'' is received, we attempt to repair the utterance. At the user's discretion, or in case of ambiguity when the phrase has multiple plausible alternatives, we can either visually or verbally prompt the user to choose from a set of actionable phrases, e.g., ``Did you mean \textlangle a\textrangle open file foo.java, \textlangle b\textrangle open folder foo/java, or \textlangle c\textrangle something else?''

To correct recognition errors, we apply Considine et al.'s~\cite{considine2022tidyparse} work on Tidyparse, which supports recognition and parsing of context-free and mildly context-sensitive grammars, and computing language edit distance. Tidyparse implements a novel approach to error correction based on the theory of context-free language reachability, conjunctive grammars and Levenshtein automata. We use a SAT solver to find the smallest edit transforming a string outside the language to a string inside the language. Only when the utterance is one or two tokens away from a known command do we attempt a repair.

VoskAPI is also capable of returning a list of alternate utterances alongside a confidence score for each. We use this list to determine if any recognized utterance is sufficiently close to an actionable command in the Idiolect grammar.

\subsection{Build Reproducibility}

One of the challenges of managing a multiplatform audio project is ensuring reproducible builds. The IntelliJ Platform does not allow shipping OS-specific plugins, which requires building a single artifact. In the past, updates would result in breaking changes on Windows, macOS, or Linux, which are difficult to debug or test from a development environment and only identified by user acceptance testing on those platforms.

To streamline the development process and mitigate the risk of platform-specific regressions in production, we set up an automated pipeline for testing and deploying the plugin. We use GitHub Actions to build the plugin for each release, and when a new commit is tagged with a release number and merged, a changelog is automatically generated from the intervening commit messages, then the plugin is signed and automatically uploaded to the JetBrains plugin repository. This allows us to quickly iterate on the plugin, test the plugin on multiple platforms, and to release new versions with assurance.

\section{Evaluation}

We conducted a preliminary experiment to evaluate recognition performance on synthetic speech. Using a set of English TTS voices (both male and female) included in macOS Ventura 13.1 with IntelliJ IDEA 2023.1, VoskAPI 0.3.45 and Idiolect 1.3.3, we synthesized 100 utterances from the predefined command list using the \texttt{say} command, then transcoded a 16000 kHz PCM signed 16-bit little-endian audio recording using \texttt{ffmpeg}, and finally transcribe the audio using VoskAPI to simulate a user-generated utterance. We report the average word error rate (WER) of three ASR models trained on US English speakers: \texttt{vosk-model-en-us-0.22} (0.63), \texttt{vosk-model-en-us-0.22-lgraph} (0.66) and \texttt{vosk-model-small-en-us-0.15} (0.73). Results on a per-voice and per-model basis are shown in  Fig.~\ref{fig:fig1} below.

\vspace{-0.4cm}
\begin{figure}[ht!]
    \centering
        \resizebox{.45\textwidth}{!}{
        \includegraphics{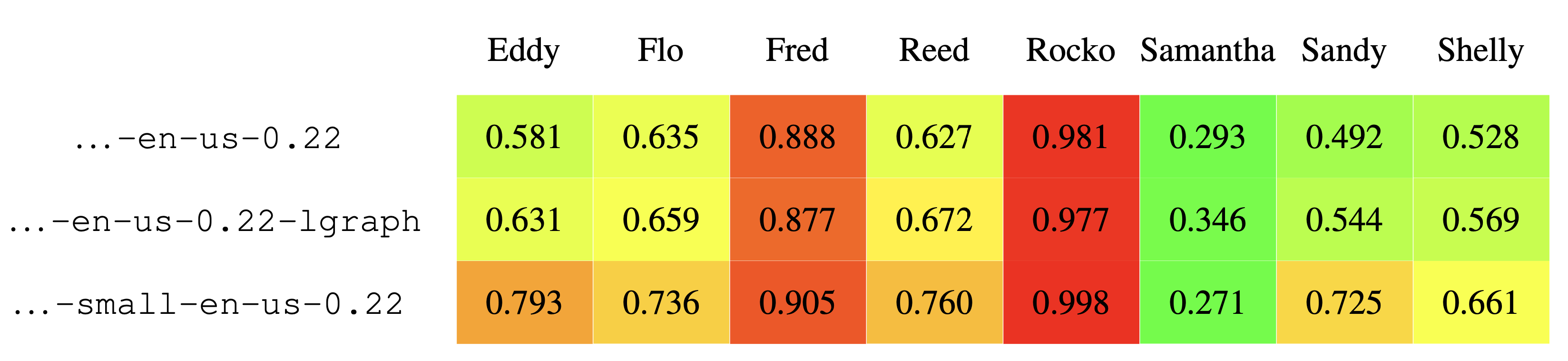}
        }
    \caption{WER for synthetic ASR using 100 commands from the default lexicon across three US English models and seven TTS voices (lower is better).}
    \label{fig:fig1}
\end{figure}

WER is a common metric for evaluating ASR performance. We use it to estimate overall intelligibility, but acknowledge it does not account for the quality or diversity of natural speech.

Two primary trends are evident. First, model size has a weak, but observable effect on recognition performance, consistent with prior work on large-vocabulary continuous speech recognition. Based on plentiful evidence in the scaling literature~\cite{droppo2021scaling}, we expect this trend to hold for larger models across natural languages. Out of the three models we tested, \texttt{vosk-model-small-en-us-0.22} exhibits the lowest average WER across all voices. Second, we note the presence of gender bias in the MacOS/Vosk TTS/ASR pipeline. Despite the apparent intelligibility of both sets of voices, WER is substantially lower on female voices (Flo, et al.) than their corresponding male counterparts (Eddy, et al.). The cause of this discrepancy is unclear, but merits further investigation.

We also evaluated the plugin using a variety of downstream metrics, such as user downloads of the plugin over a five-year timespan, bug reports, GitHub Insights, and feedback from the JetBrains Plugin Marketplace to help determine user satisfaction and identify potential areas for improvement.

\begin{figure}[ht!]
    \centering
    \includegraphics[width=0.45\textwidth]{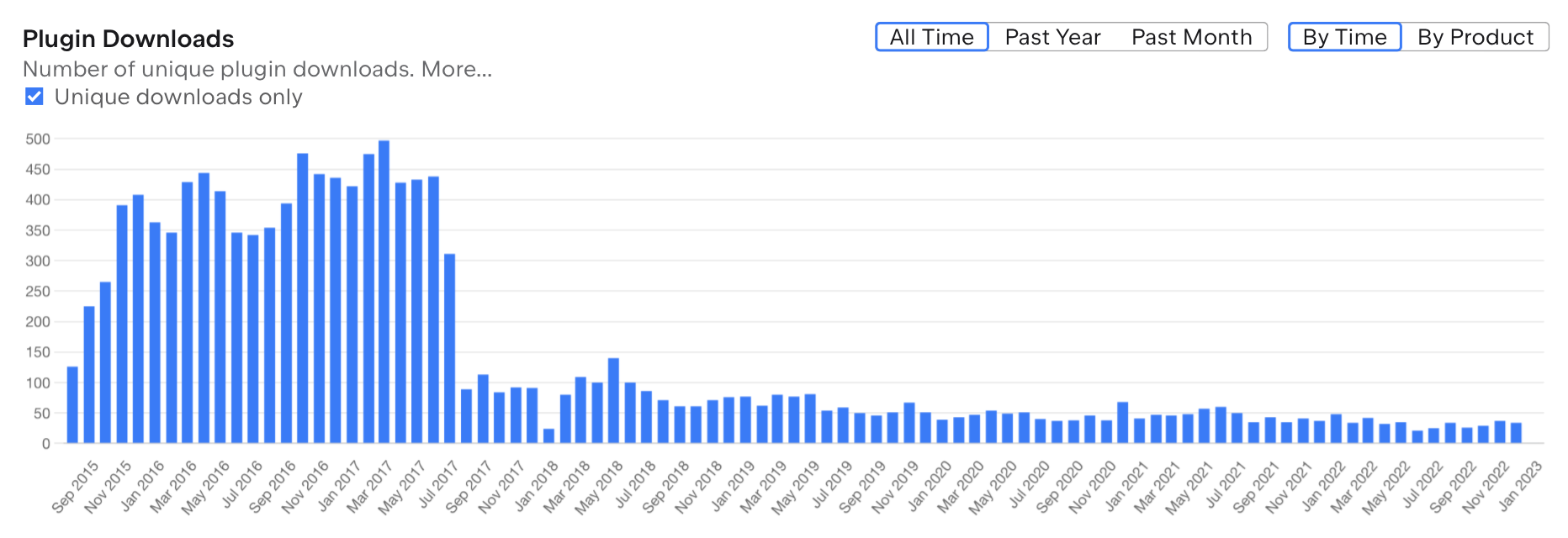}
    \caption{Plugin downloads stalled circa 2017 due to IDE incompatibility.}
\end{figure}

Downloads of the plugin stalled since its initial release in 2015. Originally based on the CMUSphinx library that was later abandoned, our plugin was downloaded over 15,964 times and 56 total bugs were filed, 14 of which were reported by external users. Following a five-year hiatus, the authors performed an intervention, rewrote the plugin from scratch based on a new speech recognition engine, implemented the usability improvements described in Sec.~\ref{sec:usability}, and republished the plugin under its current name. Our assessment is ongoing.



\section{Threats to Validity}

While voice programming may serve a niche purpose for developers, it has not been fully validated as a general programming mechanism. Our experience suggests frequent vocalizations may disturb programmers within earshot and we anticipate its usefulness in colocated workspaces will thus be limited, but do foresee some genuine use-cases, e.g., providing assistance to programmers with visual and motor disabilities.

We also acknowledge that our system is limited by its own natural language understanding capabilities. Speech is a comparatively inefficient medium for conducting programming tasks and users may be hesitant to invest time reconfiguring the system if intent recognition accuracy is low. Furthermore, the IDE plugin is currently limited to the set of actions that can be easily expressed using the IntelliJ Platform API.

Finally, we acknowledge that synthetic voices are, while increasingly lifelike, an imperfect proxy for human speech. The voices available on macOS are not a representative sample of human dialects and accents, nor we have not yet conducted a systematic human evaluation. The lexicon used to generate synthetic speech is also limited to the set of commands supported by the IDE, and does not capture how the plugin would have performed in a more general-purpose setting.

\section{Future Work}

In the future, we plan to conduct a thorough user study to better understand the use cases for the plugin. In particular, we hope to understand the development habits of visually and motor-impaired users to improve accessibility, as well as broader support for other natural languages and dialects.

The current evaluation is far from comprehensive and we intend to conduct a more thorough human evaluation of the plugin's intent-recognition capabilities to better understand the tradeoffs between accuracy, intelligibility and latency. We also hope to improve support for personalization features, such as speaker adaptation and user-specific language models.

We also aim to improve the interaction mechanism by adding audiovisual feedback and modal dialog commands. This functionality could be supported by a more user-friendly configuration interface, using an embedded DSL for defining dialog trees with a choose-your-own-adventure style modal dialog navigator. Incorporating modal logic \`a la Hazel~\cite{omar2021filling}, by adding a visual backpack and typing constraints would be an interesting direction to take and one we hope to explore.

Currently, we only track download statistics, although to inform the development of these features and better address the needs of voice programmers, it would be helpful to collect minimal telemetry on user utterances and custom commands.

Recent progress in machine learning has enabled the use of large language models (LLMs) for a variety of tasks, including speech recognition, machine translation, and text generation. An LLM trained on a large corpus of natural language can be used to perform simple reasoning tasks, such as predicting the most likely intent from a given utterance. For example, the utterance, ``I want to edit foo.java'' is more likely to trigger, ``open foo.java'' than the command, ``execute foo.java''. While LLMs are currently served on the cloud, we predict these models will become available on the edge over the next few years, enabling real-time offline speech-to-intent recognition.

Given their increasing reasoning capabilities, we would like to further explore the use of LLMs within the intent recognition pipeline. In particular, we could use language models to rerank the most likely intent corresponding to a user's utterance, conditioned on a previous historical commands, the UI state and other application-specific settings. Contextual-awareness would require more careful IDE integration and enable us to more accurately predict user intent.

\section{Conclusion}

In this work, we presented Idiolect, a plugin for the IntelliJ Platform that allows users to control the IDE and reconfigure the plugin using voice commands. This addresses a common usability issue in voice UX design, where users are unable to express their intent in a manner the system understands. We described the plugin's design and challenges we faced while implementing it, presented the results of a preliminary evaluation, and discussed avenues for future work.

On one end of the design spectrum are metaprogramming languages and software language engineering frameworks that allow users to define their own commands and build embedded domain specific languages (eDSLs). These systems can be powerful, but require a high upfront investment from the user, who must design a language, and then learn the language itself, all whilst doing their daily job as a software engineer. On the other end are finite languages, which are too inflexible.

In the middle of this language design spectrum however are \textit{idiolects}, which give users the freedom to create conversational domain-specific languages according to their own idiomatic style: this is the design space which Idiolect occupies. By targeting IDEs, whose users are typically adept programmers, commands with more complex semantics can be expressed programmatically, then invoked on-the-fly, sans recompilation.

We believe the intersection between conversational agents, vernacular programming languages and programmable voice assistants to be fertile ground, and one that is relatively unexplored in the language design space. We hope our small contribution inspires others to explore these ideas and take steps towards making programming more fun and accessible.

\section{Acknowledgements}

The first author wishes to thank his former colleagues Alexey Kudinkin and Yaroslav Lepenkin, who contributed to the plugin during the original JetBrains hackathon in 2015, and former manager Hadi Hariri for his advice and encouragement.

\bibliography{botse}
\end{document}